\documentclass[%
reprint,
showpacs,
groupedaddress,
amsmath,amssymb,
aps,
prb,
]{revtex4-1}

\usepackage{graphicx}% include figure files
\usepackage{bm}% bold math

\begin{document}

\title{Double-layered monopolar order in Tb$_2$Ti$_2$O$_7$ spin liquid}
\author{A. P. Sazonov}
\email{mail@sazonov.org}
\homepage{www.sazonov.org}
\author{A. Gukasov}
\author{I. Mirebeau}
\affiliation{CEA, Centre de Saclay, DSM/IRAMIS/Laboratoire L{\'e}on Brillouin, F-91191 Gif-sur-Yvette, France}
\author{P. Bonville}
\affiliation{CEA, Centre de Saclay, DSM/IRAMIS/Service de Physique de l'Etat Condens{\'e}, F-91191 Gif-Sur-Yvette, France}

\date{\today}% It is always \today, today, but any date may be explicitly specified

%%%%%%%%%%%%%%%%%%%%%%%%%%%%%%%%%%%%%%%%%%%%%%%%%%%%%%%%%%%%%%%%%%%%%%%%%%%%%%%
%%%%%%%%%%%%%%%%%%%%%%%%%%%%%%%%%%%%%%%%%%%%%%%%%%%%%%%%%%%%%%%%%%%%%%%%%%%%%%%
%%%%%%%%%%%%%%%%%%%%%%%%%%%%%%%%%%%%%%%%%%%%%%%%%%%%%%%%%%%%%%%%%%%%%%%%%%%%%%%

\begin{abstract}
Ho$_2$Ti$_2$O$_7$ and Dy$_2$Ti$_2$O$_7$ spin ices exhibit elementary excitations akin to magnetic monopoles. Here we focus on Tb$_2$Ti$_2$O$_7$ spin liquid, where correlated magnetic moments keep fluctuating down to very low temperatures. Using a monopole picture, we have re-analyzed the field-induced magnetic structure previously determined by neutron diffraction in Tb$_2$Ti$_2$O$_7$. We show that under a high field applied along a [110] direction, Tb$_2$Ti$_2$O$_7$ orders as a three dimensional arrangement of monopole and antimonopole double layers. In contrast, Ho$_2$Ti$_2$O$_7$ spin ice in the same conditions behaves as a monopole-free state. By symmetry analysis we derived the distortions compatible with the observed magnetic structure of Tb$_2$Ti$_2$O$_7$ which can be related to the appearance of the double-layered monopolar order.
\end{abstract}

% PACS, the Physics and Astronomy Classification Scheme.
% 71.27.+a - Strongly correlated electron systems
% 61.05.fm - Neutron diffraction
% 75.25.-j - Spin arrangements in magnetically ordered materials

%\keywords{...} % Use showkeys class option if keyword display desired

\maketitle

%%%%%%%%%%%%%%%%%%%%%%%%%%%%%%%%%%%%%%%%%%%%%%%%%%%%%%%%%%%%%%%%%%%%%%%%%%%%%%%
%%%%%%%%%%%%%%%%%%%%%%%%%%%%%%%%%%%%%%%%%%%%%%%%%%%%%%%%%%%%%%%%%%%%%%%%%%%%%%%
%%%%%%%%%%%%%%%%%%%%%%%%%%%%%%%%%%%%%%%%%%%%%%%%%%%%%%%%%%%%%%%%%%%%%%%%%%%%%%%

\section{Introduction}

\begin{figure}[b]
\includegraphics[width=1.0\columnwidth]{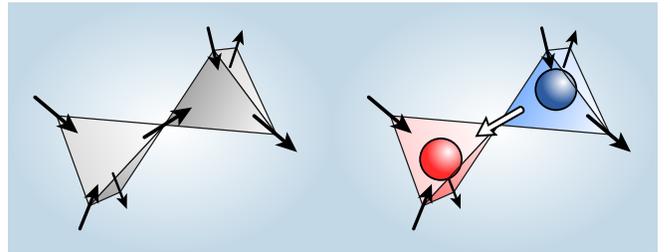}
\caption{\label{f:monopol}(Color online) Magnetic monopoles in spin ice. Left panel: Spins of two adjacent tetrahedra obeying the ``two-in, two-out'' ice rule. Right panel: Monopole--antimonopole pair induced by flipping of the spin connecting the tetrahedra.}
\end{figure}

Spin ices are geometrically frustrated magnets, which support exotic ground states and excitations~\cite{prl.81.4496.1998}. Their zero-field magnetic ground state has an extensive degeneracy issued from topological constraints, like those which govern the two energetically equivalent positions of protons of water molecules in real ice. Spin-ice behavior has been found in the pyrochlore compounds Ho$_2$Ti$_2$O$_7$ and Dy$_2$Ti$_2$O$_7$, where the rare-earth ($R$) ions reside on a lattice of corner sharing tetrahedra, and the magnetic moments behave as Ising spins, being constrained to lie along their local $\left\langle 111 \right\rangle$ anisotropy axes~\cite{sc.294.1495.2001,rmp.82.53.2010}. The $R$ magnetic moments interact via dipolar and superexchange interactions, yielding an effective nearest-neighbor ($nn$) interaction of ferromagnetic (FM) nature which governs most of the physics. In spin ices, the local structure which minimizes the energy of a single tetrahedron obeys the so-called ice rules, having two magnetic moments pointing inwards and two outwards of the tetrahedron (Fig.~\ref{f:monopol}, left panel). This situation gives zero net magnetic charge on the center of each tetrahedron. Starting from this local ``2-in, 2-out'' structure, emergent excitations can be created by flipping one magnetic moment (Fig.~\ref{f:monopol}, right panel). These point defects with ``1-in, 3-out''/``3-in, 1-out'' configuration of one tetrahedron can be considered as magnetic charges~\cite{jetp.101.481.2005} or monopoles~\cite{nt.451.42.2008}.  A spin flip induces two oppositely charged monopoles connected by a Dirac string~\cite{nt.451.42.2008} when taken apart. Because of the extensive degeneracy of the spin-ice ground state, it takes only a finite energy to separate these monopoles to infinity~\cite{nt.451.42.2008}. The energy cost necessary to violate the ice rules  and create a monopole pair can be provided  either by thermal fluctuations or by applying a magnetic field~\cite{sc.326.411.2009,*sc.326.415.2009}.

Tb$_2$Ti$_2$O$_7$ spin liquid has a different zero-field ground state, akin to a cooperative paramagnet where strongly correlated magnetic moments fluctuate at short time scales down to very low temperatures (50\,mK)~\cite{prl.82.1012.1999}. The exact reason why it behaves as a spin liquid is still a matter of debate. Virtual crystal field excitations have been previously invoked~\cite{prl.98.157204.2007,arxiv.malavian.2009}. Alternatively, we recently proposed to take into account the influence of a distortion, which induces a two singlet ground state~\cite{prb.84.184409.2011}. In any case, two key differences between Tb$_2$Ti$_2$O$_7$ and Ho/Dy spin ices must be outlined: (i) The first excited crystal field doublet lies at energy above the ground state doublet which is one order of magnitude smaller than in the spin ices, giving the magnetic moments a certain degree of freedom with respect to the local $\left\langle 111 \right\rangle$ axes; (ii) The balance between exchange and dipolar energies turns the effective $nn$ interaction to be antiferromagnetic (AFM).

In the present paper we study the long-range ordered magnetic structure which is induced in Tb$_2$Ti$_2$O$_7$ under a field $H$ applied along [110]. We have re-analyzed the magnetic structure previously determined by neutron diffraction~\cite{prl.101.196402.2008,prb.82.174406.2010} using the monopole description. We show that above a critical field the latter magnetic structure which violates the ice rules can be viewed as a three-dimensional (3D) periodic arrangement of monopole (``1-in, 3-out'') and antimonopole (``3-in, 1-out'') double layers. This structure which is unique for a given field relieves the spin-liquid degeneracy. In contrast, the spin ice Ho$_2$Ti$_2$O$_7$ in the same conditions selects a ``2-in, 2-out'' non-degenerate ground state which complies with the ice rules and can be considered as a ``vacuum'' state, namely a state without monopoles. Such description has interesting consequences when considering the spin excitations in the two compounds. A single spin flip creates either a vacancy pair (in the Tb case) or a monopole pair (in the Ho case). Whereas dipolar interactions seem to be sufficient to describe the monopole-free state of Ho$_2$Ti$_2$O$_7$ spin ice~\cite{prl.95.097202.2005}, the  double-layered monopolar state of Tb$_2$Ti$_2$O$_7$ spin liquid cannot be explained by this mechanism. We propose that it could be stabilized by a distortion, either of Jahn-Teller or spin-Peierls type.

%%%%%%%%%%%%%%%%%%%%%%%%%%%%%%%%%%%%%%%%%%%%%%%%%%%%%%%%%%%%%%%%%%%%%%%%%%%%%%%
%%%%%%%%%%%%%%%%%%%%%%%%%%%%%%%%%%%%%%%%%%%%%%%%%%%%%%%%%%%%%%%%%%%%%%%%%%%%%%%
%%%%%%%%%%%%%%%%%%%%%%%%%%%%%%%%%%%%%%%%%%%%%%%%%%%%%%%%%%%%%%%%%%%%%%%%%%%%%%%

\section{Experiment}

Single crystals of Tb$_2$Ti$_2$O$_7$ and Ho$_2$Ti$_2$O$_7$ were grown by the floating-zone technique (see, e.g., Ref.~\onlinecite{jpcm.10.L723.1998}). Neutron diffraction studies were performed on the diffractometer 5C1~($\lambda = 0.845$\,\AA) at the Orph{\'e}e reactor of the Laboratoire L{\'e}on Brillouin, Saclay. We used unpolarized neutrons, collecting typically from 200 to 400 reflections up to $\ensuremath{\sin\theta/\lambda} \approx 0.55$\,\AA$^{-1}$ for each data set at $T=1.6$\,K under the field $H \leq 7$\,T. The nuclear structure parameters were deduced from the low-temperature measurements in zero field. The integrated intensities of these reflections were used to refine the components of the Tb$^{3+}$/Ho$^{3+}$ magnetic moments using the symmetry constraints described below. The program \textsc{FullProf}~\cite{phb.192.55.1993} was used to refine the nuclear and magnetic structure.

%%%%%%%%%%%%%%%%%%%%%%%%%%%%%%%%%%%%%%%%%%%%%%%%%%%%%%%%%%%%%%%%%%%%%%%%%%%%%%%
%%%%%%%%%%%%%%%%%%%%%%%%%%%%%%%%%%%%%%%%%%%%%%%%%%%%%%%%%%%%%%%%%%%%%%%%%%%%%%%
%%%%%%%%%%%%%%%%%%%%%%%%%%%%%%%%%%%%%%%%%%%%%%%%%%%%%%%%%%%%%%%%%%%%%%%%%%%%%%%

\section{Magnetic structure analysis}

\begin{figure*}
\includegraphics[width=1.0\textwidth]{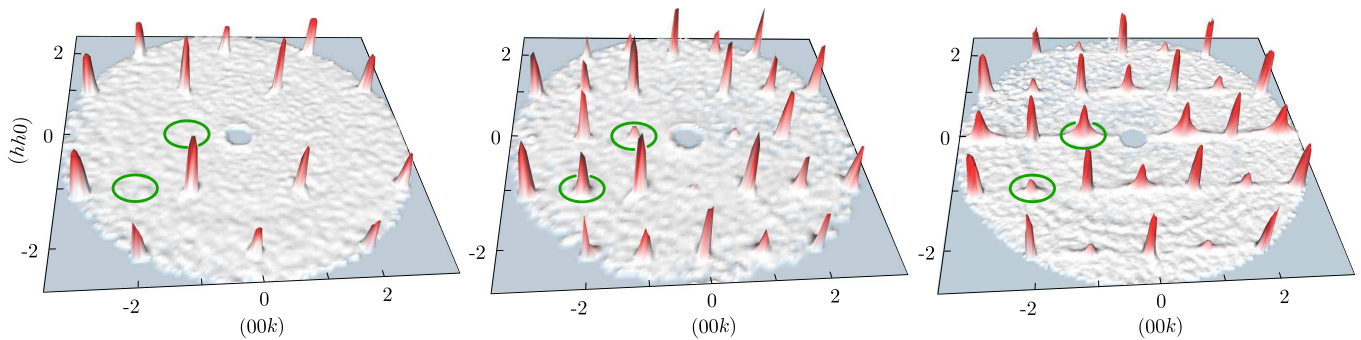}
\caption{\label{f:neutron-scat}(Color online) Neutron scattering of Tb$_2$Ti$_2$O$_7$ in 0\,T (left panel), Tb$_2$Ti$_2$O$_7$ in 5\,T (middle panel) and Ho$_2$Ti$_2$O$_7$ in 2\,T (right panel) at 1.6\,K with $\bm{H} \parallel [110]$. Intensity is given in a logarithmic scale. Selected magnetic reflections are highlighted. In zero field, only the peaks of the crystal structure are seen.}
\end{figure*}

In rare-earth pyrochlores, the magnetic structure induced by a field $\bm{H} \parallel [110]$ splits into orthogonal $R$-$\alpha$ and $R$-$\beta$ chains, oriented along [110] and [1$\bar{1}$0] respectively. $R$ ions on the $\alpha$ chains have their local anisotropy axis at $\pm 35.26$$^\circ$ from the field, whereas in the $\beta$ chains the local anisotropy axes are perpendicular to $\bm{H}$. In Tb$_2$Ti$_2$O$_7$, when the field increases along [110], the ordering of the Tb moments occurs in two steps~\cite{prl.96.177201.2006,prb.82.100401.2010,prb.82.174406.2010}. In the low-field region,  magnetic order with $\bm{k} = \mathbf{0}$ propagation vector is induced. The magnetic moments in the FM-like $\alpha$ chains increase linearly with $H$ and tend to saturate ($\approx 6$\,$\mu_\text{B}$) in a 1\,T field. The FM-like order in the $\alpha$ chains means that the chains have non-zero homogeneous magnetic component parallel to the field. The moments induced on the $\beta$ chains remain quite small since the field is applied along the local hard axes of the Tb moments. Above a critical field ($H_\text{C} \approx 2$\,T at $T = 50$\,mK), the AFM order both inside and between the $\beta$ chains occur. This is shown by the appearance of an additional $\bm{k} = (0,0,1)$ magnetic structure which yields magnetic intensity on the Bragg peaks of the simple cubic lattice (Fig.~\ref{f:neutron-scat}, middle panel), whereas below 2\,T the magnetic intensity is restricted to the peaks of the face-centered cubic lattice (Fig.~\ref{f:neutron-scat}, left panel), in agreement with the $Fd\bar{3}m$ space group of the pyrochlore structure. The $\bm{k} = (0,0,1)$ structure crucially depends on the temperature $T$, the field $H$ and its precise orientation $\Delta\phi$ with respect to the [110] axis~\cite{prb.82.100401.2010,prb.82.174406.2010}.

To describe these structures, we previously performed a symmetry analysis based on the theory of representations and found that both $\bm{k} = \mathbf{0}$ and $\bm{k} = (0,0,1)$ magnetic structures can be described by single irreducible representations~\cite{prb.82.174406.2010,jpcm.23.164221.2011}. This reduces the number of the refined parameters from 12 in an unconstrained refinement (3 for each of the 4 Tb/Ho moments in a tetrahedron), to 6 and 4 in case of the $\bm{k} = \mathbf{0}$ and $\bm{k} = (0,0,1)$ structures, respectively. A detailed description of the analysis can be found in Ref.~\onlinecite{jpcm.23.164221.2011}.

\begin{figure}
\includegraphics[width=1.0\columnwidth]{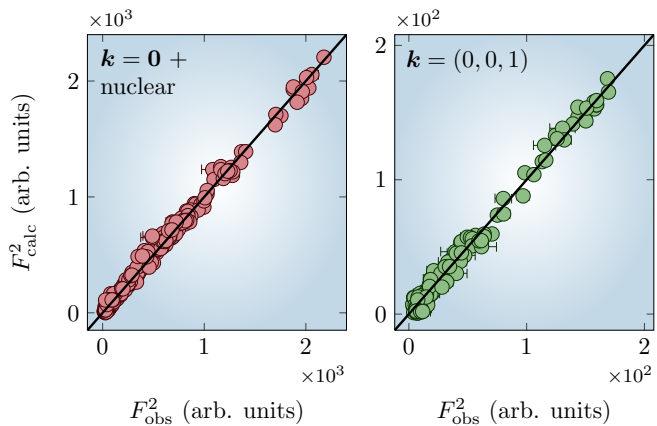}
\caption{\label{f:refinement}(Color online) A typical fit for Tb$_2$Ti$_2$O$_7$ at 1.6\,K and 7\,T. The calculated integrated intensities ($F^2_\text{calc}$) are plotted against the experimental ones ($F^2_\text{obs}$). Left panel: Both magnetic $\bm{k} = \mathbf{0}$ and nuclear intensities. Right panel: Pure magnetic intensities of the $\bm{k} = (0,0,1)$ structure.}
\end{figure}

An example of a typical fit for both $\bm{k} = \mathbf{0}$ and $\bm{k} = (0,0,1)$ structures of Tb$_2$Ti$_2$O$_7$ at 1.6\,K and 7\,T is shown in Fig.~\ref{f:refinement}. The resulting magnetic structure of Tb$_2$Ti$_2$O$_7$ at 1.6\,K in 7\,T calculated as a superposition of the $\bm{k} = \mathbf{0}$ and (0,0,1) structures is shown in Fig.~\ref{f:magn-str}. On each tetrahedron, both $\beta$ spins point either ``in'' or ``out'' (AFM behavior) and the $\alpha$ chain is made of alternating moments (FM-like behavior), so that the local spin configuration (``1-in, 3-out''/``3-in, 1-out'') violates the ice rules.

The high-field magnetic structure of Tb$_2$Ti$_2$O$_7$ deduced from this symmetry analysis agrees with our earlier results~\cite{prl.101.196402.2008} and is consistent with the schematic picture proposed in Ref.~\onlinecite{prb.82.100401.2010} (see Fig.~1) for Ising spins. We notice however that the  model proposed in Ref.~\onlinecite{prb.82.100401.2010} gives a magnetization value about three times smaller than that reported in the literature~\cite{jpsj.71.599.2002}. This discrepancy is likely to be attributed to strong extinction corrections which probably affected the intensities of 3 Bragg reflections measured with cold incident neutrons (5\,\AA)~\cite{prb.82.100401.2010}. Our data analysis involves 200 to 400 Bragg reflections measured with hot neutrons (0.845\,\AA) and weakly influenced by extinction. This lifts the ambiguities of Ref.~\onlinecite{prb.82.100401.2010} and all our results perfectly agree with the magnetization measurements~\cite{jpsj.71.599.2002}. We also notice that in the real magnetic structure deduced from our analysis, there is a small deviation of the Tb magnetic moments from the local Ising $\left\langle 111 \right\rangle$ axes, which is discussed below.

%%%%%%%%%%%%%%%%%%%%%%%%%%%%%%%%%%%%%%%%%%%%%%%%%%%%%%%%%%%%%%%%%%%%%%%%%%%%%%%
%%%%%%%%%%%%%%%%%%%%%%%%%%%%%%%%%%%%%%%%%%%%%%%%%%%%%%%%%%%%%%%%%%%%%%%%%%%%%%%
%%%%%%%%%%%%%%%%%%%%%%%%%%%%%%%%%%%%%%%%%%%%%%%%%%%%%%%%%%%%%%%%%%%%%%%%%%%%%%%

\section{Monopole picture}

The high-field magnetic structure of Tb$_2$Ti$_2$O$_7$ can be viewed as tetrahedra layers of ``1-in, 3-out'' type alternating with layers of opposite type ``3-in, 1-out''. Such local configurations are exactly the defects with respect to the ice rules, namely the monopoles, which are the building bricks of the magnetic monopole representation in spin ices~\cite{nt.451.42.2008}. Thus, the Tb$_2$Ti$_2$O$_7$ magnetic structure consists of a 3D array of alternating double layers of monopoles and antimonopoles (see Fig.~\ref{f:excitation}), corresponding to positive and negative magnetic charges, and stacked perpendicular to [001]. Note that this periodicity is different from that of the staggered monopoles structure~\cite{jpsj.78.103706.2009} observed in spin ices when the field is along the [111] direction.

\begin{figure}
\includegraphics[width=1.0\columnwidth]{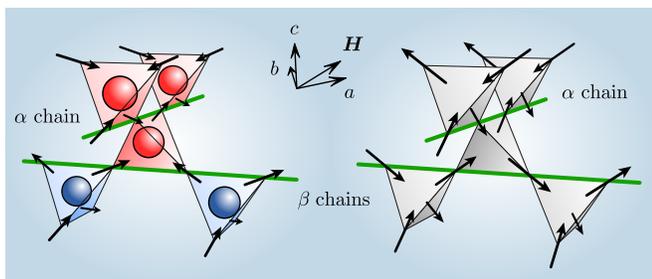}
\caption{\label{f:magn-str}(Color online) Magnetic structures of Tb$_2$Ti$_2$O$_7$ (left panel) and Ho$_2$Ti$_2$O$_7$ (right panel) in a [110] field. Blue and red balls correspond to opposite magnetic charges. The orientations of the $\beta$ moments inside a chain are AFM in Tb$_2$Ti$_2$O$_7$ and FM-like in Ho$_2$Ti$_2$O$_7$.}
\end{figure}

\begin{figure*}
\includegraphics[width=1.0\textwidth]{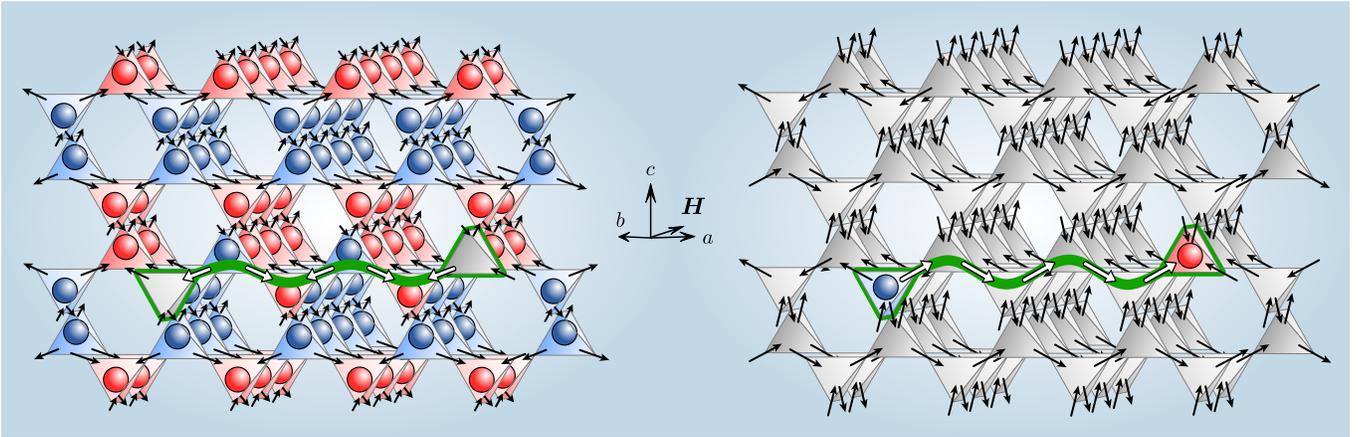}
\caption{\label{f:excitation}(Color online) Left panel: Double-layered monopolar structure of Tb$_2$Ti$_2$O$_7$ with vacuum pair excitations. Right panel: Magnetically vacuum state of Ho$_2$Ti$_2$O$_7$ with monopole pair excitation.}
\end{figure*}

In contrast, in the spin ice Ho$_2$Ti$_2$O$_7$ studied in the same conditions, the highest intensity peaks measured in neutron diffraction have the lowest intensity in Tb$_2$Ti$_2$O$_7$ and vice versa (Fig.~\ref{f:neutron-scat}, right panel). The same symmetry analysis shows that in Ho$_2$Ti$_2$O$_7$ the $\beta$ chains  also order antiferromagnetically with respect to each other above $H \approx 0.1$\,T. But here the orientation of the $\beta$ moments alternating ``in'' and ``out'' combined with the $\alpha$ moments alternating in the same way preserve the ice rules in this single state (Fig.~\ref{f:magn-str}, right panel)~\cite{prl.101.196402.2008,prb.82.174406.2010}. The refined magnetic structure is close to the schematic $X$ structure~\cite{prl.79.2554.1997}, with the difference that the moments of the $\beta$ chains are lower (8\,$\mu_\text{B}$) than the saturation value of 10\,$\mu_\text{B}$~\cite{jpcm.23.164221.2011}. We recall that the $X$ structure involves $\alpha$ and $\beta$ moments of equal magnitudes aligned along the local $\left\langle 111 \right\rangle$ axes, but alternating to keep a net ferromagnetic moment in a given chain (FM-like behavior). In this  single state, the antiferromagnetic coupling of the $\beta$ chains together with the ferromagnetic coupling of the $\alpha$ chains preserves the ice rules. The reduced moment value obtained in the experiments and the Lorenzian peak shape shown in Fig.~\ref{f:neutron-scat} come from a partial disorder among the $\beta$ chains, as discussed below.

Neglecting this effect and using the monopole description as above, the ordered state of Ho$_2$Ti$_2$O$_7$ spin ice in a magnetic field $\bm{H} \parallel [110]$ contains no monopoles in the ground state, since  the monopole charges compensate in each tetrahedron. That is, the ground state of Ho$_2$Ti$_2$O$_7$ in a field is ``2-in, 2-out'' structure, which represents a ``vacuum'' state from the monopole point of view. We underline that the main difference between the field-induced magnetic structures of Tb$_2$Ti$_2$O$_7$ and Ho$_2$Ti$_2$O$_7$ comes from their $\bm{k} = (0,0,1)$ structures, namely, the type of ordering within the $\beta$ chains, which is responsible for the overall monopolar/vacuum order. These AFM $\bm{k} = (0,0,1)$ structures are described by different magnetic space groups $Inna$ and $Inma$ for Tb$_2$Ti$_2$O$_7$ and Ho$_2$Ti$_2$O$_7$, respectively, whereas both compounds have the same magnetic symmetry $Imm'a'$ of their FM-like $\bm{k} = \mathbf{0}$ structures.

Starting from these two ordered states mentioned above for both Tb$_2$Ti$_2$O$_7$ and Ho$_2$Ti$_2$O$_7$, elementary excitations can be viewed as follows. In Tb$_2$Ti$_2$O$_7$, an excitation in the double-layered monopolar state is obtained by reversing the orientation of a single magnetic moment between a monopole and an antimonopole (but not between two monopoles of the same ``type'') to recover the spin ice ``2-in, 2-out'' structure in the tetrahedra involved. This corresponds to a recombination of two monopole and antimonopole nearest neighbors, which produces a pair of magnetically neutral vacancies. Such vacancies can move away from each other (Fig.~\ref{f:excitation}, left panel) with little energy cost by further flips of adjacent magnetic moments provided they remain within the same $\beta$ chain. Hence although monopoles order in three dimensions, their excitations only propagate in one dimension along the $\beta$ chains. In turn, the $\alpha$ moments with anisotropy axes close to the field experience a large Zeeman energy and hence cannot flip easily. Alternatively, the propagation of a monopole pair in the vacuum state of Ho$_2$Ti$_2$O$_7$ in a [110] field (Fig.~\ref{f:excitation}, right panel) mirrors that found in Tb$_2$Ti$_2$O$_7$.

In the high-field ordered states of Tb$_2$Ti$_2$O$_7$ and Ho$_2$Ti$_2$O$_7$ with $\bm{H} \parallel [110]$, defects (namely monopoles or monopole vacancies) can be created through a small field misalignment. They can be evidenced in neutron diffraction experiments as deviation from the perfect Bragg reflection, such as a peak broadening or diffuse scattering. Increasing the field misalignment and the number of defects breaks the monopolar/vacuum order in both Tb$_2$Ti$_2$O$_7$ and Ho$_2$Ti$_2$O$_7$ compounds~\cite{jpcm.16.R1277.2004,*prb.79.014408.2009,prb.82.174406.2010}, but the mechanisms are different. Due to its peculiar crystal field scheme~\cite{prb.76.184436.2007}, Tb$^{3+}$ has a finite value of the $g_{\perp}$ component of the Land\'{e} tensor, $g_{\perp} / g_{\parallel} \approx 1/5$ in a field of 1\,T~\cite{prl.103.056402.2009}, an order of magnitude larger than those in Ho or Dy spin ices with Ising behavior. This non-negligible $g_{\perp}$ leads to the appearance of a magnetic component along the field in the $\beta$ chains, so that the magnetic moments deviate from their local $\left\langle 111 \right\rangle$ Ising axes (Fig.~\ref{f:magn-str}). Consequently, in Tb$_2$Ti$_2$O$_7$ the double-layered monopolar order disappears when increasing the field misalignment, into a spin-ice-like configuration involving moments of unequal magnitudes~\cite{prl.101.196402.2008,prb.82.174406.2010}. In contrast, in Ho$_2$Ti$_2$O$_7$ the deviation of the ordered moments with respect to the anisotropy axes is negligible. A huge precision in the field alignment is necessary to achieve a fully ordered vacuum state ($X$ structure). Such a vacuum state breaks down with the misalignment and another ordered state which also obeys the ice rules, namely the $\bm{k}=\bm{0}$ structure, is stabilized.

As shown previously~\cite{prl.95.097202.2005}, dipolar interactions are sufficient to describe the high-field state of Ho$_2$Ti$_2$O$_7$. In dipolar spin ices (DSI) where the dipolar interaction dominates over the exchange, the isolated spin chains tend to order ferromagnetically. The simplest model that captures the essential features of the behavior of the $\beta$ chains in a magnetic field is the model of non-interacting Ising chains with $nn$ exchange and dipolar interactions~\cite{jpsj.72.3045.2003}. We adapt this model to Tb$_2$Ti$_2$O$_7$, as shown below.

The magnetic interaction Hamiltonian is taken as
\begin{center}
\begin{math}
\begin{array}{ll}
  \mathcal{H} = \mathcal{H}^\text{ex} + \mathcal{H}^\text{dip} = & -\mathcal{J} \, \sum_{\left\langle i,j \right\rangle} \, \bm{J}_i \cdot \bm{J}_j \\
  & + \mathcal{D} \, R_{nn}^3 \, \sum_{i>j} \frac{\bm{J}_i \cdot \bm{J}_j - 3 (\bm{J}_i \cdot \bm{\hat{R}}_{ij}) (\bm{J}_j \cdot \bm{\hat{R}}_{ij})} {|\bm{R}_{ij}|^{3}}.
\end{array}
\end{math}
\end{center}
Here $\bm{R}_{ij} \equiv \bm{R}_{j} - \bm{R}_{i} = |\bm{R}_{ij}|\bm{\hat{R}}_{ij}$, $\bm{R}_{i}$ is the position vector of atom $i$ with total angular momentum $\bm{J}_{i}$, $R_{nn}$ is the distance between $nn$ atoms, $\mathcal{D} = \mu_0 (g_J \mu_\text{B})^2(4\pi R_{nn}^3)^{-1}$ is the dipolar coupling, $g_J$ is the Land{\'e} factor, and $\mathcal{J}$ is the $nn$ exchange coupling with $\mathcal{J} < 0$ for AFM.

In order to estimate the strength of the dipolar and exchange interactions of a single chain in high field (above 4\,T), we use our experimental values of the Tb moment orientation, which is governed by the crystal field. The Tb-$\beta$ moments of the $\bm{k} = (0,0,1)$ structure deviate from the local $\left\langle 111 \right\rangle$ axes by about 13$^\circ$. In this case, the $nn$ exchange interaction is simplified to
\begin{equation*}
  \mathcal{H}^\text{ex} \approx 0.7 \, \mathcal{J} \mu^2 (g_J \mu_\text{B})^{-2},
\end{equation*}
where $\mu$ is the Tb magnetic moment. The long-range dipolar interactions can be calculated by considering the interaction between the atom and its odd neighbors at a distance $R_{2m+1} = (2m+1) \, R_{nn}$ or its even neighbors at a distance $R_{2m} = 2m \, R_{nn}$ as
\begin{equation*}
\begin{split}
  \mathcal{H}^\text{dip} & \approx \mathcal{D} \mu^2 (g_J \mu_\text{B})^{-2} \left[ 1.8 \sum (2m+1)^{-3} - 1.5 \sum (2m)^{-3} \right] \\
  & \approx 1.4 \, A_p \, \mathcal{D} \mu^2 (g_J \mu_\text{B})^{-2} \approx 1.7 \, \mathcal{D} \mu^2 (g_J \mu_\text{B})^{-2},
\end{split}
\end{equation*}
with Ap\'{e}ry constant $A_{p} = \sum 1/m^{3} \approx 1.2$.

Similar to dipolar spin ices~\cite{sc.294.1495.2001}, to take exchange and dipolar interactions into account in Tb$_2$Ti$_2$O$_7$, we define an effective energy scale, $\mathcal{H}_\text{eff}$, as
\begin{equation*}
  \mathcal{H}_\text{eff} = (0.7 \, \mathcal{J} + 1.7 \, \mathcal{D}) \, \mu^2 (g_J \mu_\text{B})^{-2},
\end{equation*}
yielding an effective exchange $\mathcal{J}_\text{eff} \equiv 0.7\mathcal{J} + 1.7\mathcal{D}$.

So, $\mathcal{J}_\text{eff}$ is expected to be negative (antiferromagnetic) when $\mathcal{J}/\mathcal{D} \lesssim -1.7/0.7 \approx -2.4$. We estimate the value of $\mathcal{J}/\mathcal{D}$ considering the exchange parameters for Tb$_2$Ti$_2$O$_7$ reported in the literature. Taking into account the experimental value of $R_{nn} \approx 3.59$\,\AA\ for Tb$_2$Ti$_2$O$_7$, the dipolar coupling is estimated as $\mathcal{D} \approx 0.0315$\,K~\cite{prl.98.157204.2007}. There are two different published values of $\mathcal{J}$, namely, $-0.083$\,K~\cite{prb.76.184436.2007} and $-0.167$\,K~\cite{prl.98.157204.2007}, giving $\mathcal{J}/\mathcal{D}$ of about $-2.6$ and $-5.3$, respectively. Nevertheless, both of them yield a negative $\mathcal{J}_\text{eff}$ and are in agreement with an AFM order within the $\beta$ chain. Note that $\mathcal{J} \approx -0.083$\,K~\cite{prb.76.184436.2007} is expected to be more reliable as it takes into account the full crystal field scheme of Tb$_2$Ti$_2$O$_7$.

\begin{figure}
\includegraphics[width=1.0\columnwidth]{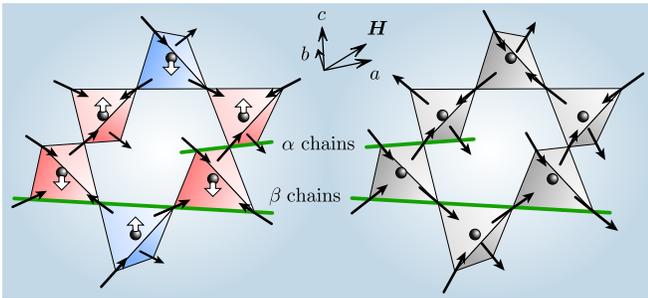}
\caption{\label{f:distortion}(Color online) Left panel: The structural distortion in Tb$_2$Ti$_2$O$_7$ compatible with the observed high-field magnetic structure. The directions of displacements of the axial oxygens are shown by white arrows. Right panel: The undistorted structure of Ho$_2$Ti$_2$O$_7$.}
\end{figure}

Thus, the comparison between the strength (and signs) of the dipolar and exchange interactions is consistent with the AFM order inside the $\beta$ chains. However, the model applied in here deals with a single chain and thus it cannot explain the full 3D magnetic order. Therefore, this oversimplified model should be extended by taking into account both the anisotropic exchange and the full crystal field scheme. Moreover, numerical Monte Carlo studies~\cite{prl.95.097202.2005} show that in the $X$ structure of the Ho$_2$Ti$_2$O$_7$ spin ice, both the FM moment components along a $\beta$ chain and the AFM orientation of $nn$ chains arise from the dipolar interaction. This explanation does not hold for Tb$_2$Ti$_2$O$_7$ where an additional mechanism should be at play. We propose to consider the influence of a distortion, of Jahn-Teller or spin-Peierls type. Structural fluctuations were observed in Tb$_2$Ti$_2$O$_7$ at low temperature~\cite{prl.99.237202.2007}, possibly precursor of a $T=0$ Jahn-Teller transition. A spontaneous Jahn-Teller distortion could be the main ingredient to explain the spin-liquid ground state of Tb$_2$Ti$_2$O$_7$ in zero field~\cite{prb.84.184409.2011}. Exchange driven spin-Peierls distortions have also been considered as possible mechanisms to relieve the degeneracy of the pyrochlore lattice, in spinels compounds especially~\cite{prl.85.4960.2000,prl.88.067203.2002}.

Up to now, the structural distortion in Tb$_2$Ti$_2$O$_7$, either spontaneous~\cite{prl.99.237202.2007}, field induced~\cite{prl.105.077203.2010} or induced by pressure/stress~\cite{prl.93.187204.2004}, has not been characterized. In our case when the field applied along the [110] axis its symmetry should reflect that of the magnetic order. We derived the distortions compatible with the observed magnetic modes by symmetry analysis. Interestingly, we found that only one distorted structure has a symmetry compatible with that of the magnetic order of Tb$_2$Ti$_2$O$_7$ in high field, and that this structure involves displacements of the oxygen ions exclusively. We recall that in the pyrochlore structure, each rare-earth ion is surrounded by eight oxygens, two ``axial'' ones along the local $\left\langle 111 \right\rangle$ axis and six ``planar'' ones lying in a buckled perpendicular plane. The symmetry constraints on the allowed displacive modes are very different for axial and planar oxygens. The displacement of axial oxygen is allowed only along the [001] axis and occurs in such a way that half of the oxygens move parallel and half antiparallel to this axis (see Fig.~\ref{f:distortion}, left panel). The displacive mode of planar oxygens is more complex but can be also explicited. However, we believe that the displacements of planar oxygens have less effect on the magnetic interactions and on the crystal field of rare-earth pyrochlores, since these oxygens are typically about 13\,\%  farther from rare-earth ions than the axial ones. Hence, we suggest that in the monopole picture proposed here for Tb$_2$Ti$_2$O$_7$ in high magnetic field along [110], oxygen displacement should result from the attraction (repulsion) of monopoles with opposite (identical) magnetic charges. For comparison,  we also show the reference structure of Ho$_2$Ti$_2$O$_7$ (see Fig.~\ref{f:distortion}, right panel) where such distortions are expected to be negligibly small, due to its different crystal field scheme.

%%%%%%%%%%%%%%%%%%%%%%%%%%%%%%%%%%%%%%%%%%%%%%%%%%%%%%%%%%%%%%%%%%%%%%%%%%%%%%%
%%%%%%%%%%%%%%%%%%%%%%%%%%%%%%%%%%%%%%%%%%%%%%%%%%%%%%%%%%%%%%%%%%%%%%%%%%%%%%%
%%%%%%%%%%%%%%%%%%%%%%%%%%%%%%%%%%%%%%%%%%%%%%%%%%%%%%%%%%%%%%%%%%%%%%%%%%%%%%%

\section{Conclusion}

In conclusion, we show that under a high magnetic field applied along a [110] direction, the magnetic structure of Tb$_2$Ti$_2$O$_7$ can be viewed as a 3D arrangement of monopole and antimonopole double layers, in contrast to Ho$_2$Ti$_2$O$_7$ where the high-field magnetic structure behaves as a vacuum (monopole-free) state. A toy model for an isolated chain shows that in Tb$_2$Ti$_2$O$_7$ the exchange interaction overcomes the dipolar one, which is consistent with AFM order in the $\beta$ chains, but understanding the full 3D magnetic structure of Tb$_2$Ti$_2$O$_7$ requires to take into account the real crystal field, anisotropic exchange, dipolar interactions and the influence of a distortion. Finally, symmetry analysis allows us to derive a distortion model compatible with the observed magnetic structure of Tb$_2$Ti$_2$O$_7$ which can be related to the appearance of the double-layered monopolar state.

%%%%%%%%%%%%%%%%%%%%%%%%%%%%%%%%%%%%%%%%%%%%%%%%%%%%%%%%%%%%%%%%%%%%%%%%%%%%%%%
%%%%%%%%%%%%%%%%%%%%%%%%%%%%%%%%%%%%%%%%%%%%%%%%%%%%%%%%%%%%%%%%%%%%%%%%%%%%%%%
%%%%%%%%%%%%%%%%%%%%%%%%%%%%%%%%%%%%%%%%%%%%%%%%%%%%%%%%%%%%%%%%%%%%%%%%%%%%%%%
%\bibliography{biblio_short-9}

\begin{thebibliography}{33}%
\makeatletter
\providecommand \@ifxundefined [1]{%
 \@ifx{#1\undefined}
}%
\providecommand \@ifnum [1]{%
 \ifnum #1\expandafter \@firstoftwo
 \else \expandafter \@secondoftwo
 \fi
}%
\providecommand \@ifx [1]{%
 \ifx #1\expandafter \@firstoftwo
 \else \expandafter \@secondoftwo
 \fi
}%
\providecommand \natexlab [1]{#1}%
\providecommand \enquote  [1]{``#1''}%
\providecommand \bibnamefont  [1]{#1}%
\providecommand \bibfnamefont [1]{#1}%
\providecommand \citenamefont [1]{#1}%
\providecommand \href@noop [0]{\@secondoftwo}%
\providecommand \href [0]{\begingroup \@sanitize@url \@href}%
\providecommand \@href[1]{\@@startlink{#1}\@@href}%
\providecommand \@@href[1]{\endgroup#1\@@endlink}%
\providecommand \@sanitize@url [0]{\catcode `\\12\catcode `\$12\catcode
  `\&12\catcode `\#12\catcode `\^12\catcode `\_12\catcode `\%12\relax}%
\providecommand \@@startlink[1]{}%
\providecommand \@@endlink[0]{}%
\providecommand \url  [0]{\begingroup\@sanitize@url \@url }%
\providecommand \@url [1]{\endgroup\@href {#1}{\urlprefix }}%
\providecommand \urlprefix  [0]{URL }%
\providecommand \Eprint [0]{\href }%
\providecommand \doibase [0]{http://dx.doi.org/}%
\providecommand \selectlanguage [0]{\@gobble}%
\providecommand \bibinfo  [0]{\@secondoftwo}%
\providecommand \bibfield  [0]{\@secondoftwo}%
\providecommand \translation [1]{[#1]}%
\providecommand \BibitemOpen [0]{}%
\providecommand \bibitemStop [0]{}%
\providecommand \bibitemNoStop [0]{.\EOS\space}%
\providecommand \EOS [0]{\spacefactor3000\relax}%
\providecommand \BibitemShut  [1]{\csname bibitem#1\endcsname}%
\let\auto@bib@innerbib\@empty
%</preamble>
\bibitem [{\citenamefont {Harris}\ \emph {et~al.}(1998)\citenamefont {Harris},
  \citenamefont {Bramwell}, \citenamefont {Holdsworth},\ and\ \citenamefont
  {Champion}}]{prl.81.4496.1998}%
  \BibitemOpen
  \bibfield  {author} {\bibinfo {author} {\bibfnamefont {M.~J.}\ \bibnamefont
  {Harris}}, \bibinfo {author} {\bibfnamefont {S.~T.}\ \bibnamefont
  {Bramwell}}, \bibinfo {author} {\bibfnamefont {P.~C.~W.}\ \bibnamefont
  {Holdsworth}}, \ and\ \bibinfo {author} {\bibfnamefont {J.~D.~M.}\
  \bibnamefont {Champion}},\ }\href@noop {} {\bibfield  {journal} {\bibinfo
  {journal} {Phys. Rev. Lett.}\ }\textbf {\bibinfo {volume} {81}},\ \bibinfo
  {pages} {4496} (\bibinfo {year} {1998})}\BibitemShut {NoStop}%
\bibitem [{\citenamefont {Bramwell}\ and\ \citenamefont
  {Gingras}(2001)}]{sc.294.1495.2001}%
  \BibitemOpen
  \bibfield  {author} {\bibinfo {author} {\bibfnamefont {S.~T.}\ \bibnamefont
  {Bramwell}}\ and\ \bibinfo {author} {\bibfnamefont {M.~J.~P.}\ \bibnamefont
  {Gingras}},\ }\href@noop {} {\bibfield  {journal} {\bibinfo  {journal}
  {Science}\ }\textbf {\bibinfo {volume} {294}},\ \bibinfo {pages} {1495}
  (\bibinfo {year} {2001})}\BibitemShut {NoStop}%
\bibitem [{\citenamefont {Gardner}\ \emph {et~al.}(2010)\citenamefont
  {Gardner}, \citenamefont {Gingras},\ and\ \citenamefont
  {Greedan}}]{rmp.82.53.2010}%
  \BibitemOpen
  \bibfield  {author} {\bibinfo {author} {\bibfnamefont {J.~S.}\ \bibnamefont
  {Gardner}}, \bibinfo {author} {\bibfnamefont {M.~J.~P.}\ \bibnamefont
  {Gingras}}, \ and\ \bibinfo {author} {\bibfnamefont {J.~E.}\ \bibnamefont
  {Greedan}},\ }\href@noop {} {\bibfield  {journal} {\bibinfo  {journal} {Rev.
  Mod. Phys.}\ }\textbf {\bibinfo {volume} {82}},\ \bibinfo {pages} {53}
  (\bibinfo {year} {2010})}\BibitemShut {NoStop}%
\bibitem [{\citenamefont {Ryzhkin}(2005)}]{jetp.101.481.2005}%
  \BibitemOpen
  \bibfield  {author} {\bibinfo {author} {\bibfnamefont {I.~A.}\ \bibnamefont
  {Ryzhkin}},\ }\href@noop {} {\bibfield  {journal} {\bibinfo  {journal} {J.
  Exp. Theor. Phys.}\ }\textbf {\bibinfo {volume} {101}},\ \bibinfo {pages}
  {481} (\bibinfo {year} {2005})}\BibitemShut {NoStop}%
\bibitem [{\citenamefont {Castelnovo}\ \emph {et~al.}(2008)\citenamefont
  {Castelnovo}, \citenamefont {Moessner},\ and\ \citenamefont
  {Sondhi}}]{nt.451.42.2008}%
  \BibitemOpen
  \bibfield  {author} {\bibinfo {author} {\bibfnamefont {C.}~\bibnamefont
  {Castelnovo}}, \bibinfo {author} {\bibfnamefont {R.}~\bibnamefont
  {Moessner}}, \ and\ \bibinfo {author} {\bibfnamefont {S.~L.}\ \bibnamefont
  {Sondhi}},\ }\href@noop {} {\bibfield  {journal} {\bibinfo  {journal}
  {Nature}\ }\textbf {\bibinfo {volume} {451}},\ \bibinfo {pages} {42}
  (\bibinfo {year} {2008})}\BibitemShut {NoStop}%
\bibitem [{\citenamefont {Morris}\ \emph {et~al.}(2009)\citenamefont {Morris},
  \citenamefont {Tennant}, \citenamefont {Grigera}, \citenamefont {Klemke},
  \citenamefont {Castelnovo}, \citenamefont {Moessner}, \citenamefont
  {Czternasty}, \citenamefont {Meissner}, \citenamefont {Rule}, \citenamefont
  {Hoffmann}, \citenamefont {Kiefer}, \citenamefont {Gerischer}, \citenamefont
  {Slobinsky},\ and\ \citenamefont {Perry}}]{sc.326.411.2009}%
  \BibitemOpen
  \bibfield  {author} {\bibinfo {author} {\bibfnamefont {D.~J.~P.}\
  \bibnamefont {Morris}}, \bibinfo {author} {\bibfnamefont {D.~A.}\
  \bibnamefont {Tennant}}, \bibinfo {author} {\bibfnamefont {S.~A.}\
  \bibnamefont {Grigera}}, \bibinfo {author} {\bibfnamefont {B.}~\bibnamefont
  {Klemke}}, \bibinfo {author} {\bibfnamefont {C.}~\bibnamefont {Castelnovo}},
  \bibinfo {author} {\bibfnamefont {R.}~\bibnamefont {Moessner}}, \bibinfo
  {author} {\bibfnamefont {C.}~\bibnamefont {Czternasty}}, \bibinfo {author}
  {\bibfnamefont {M.}~\bibnamefont {Meissner}}, \bibinfo {author}
  {\bibfnamefont {K.~C.}\ \bibnamefont {Rule}}, \bibinfo {author}
  {\bibfnamefont {J.-U.}\ \bibnamefont {Hoffmann}}, \bibinfo {author}
  {\bibfnamefont {K.}~\bibnamefont {Kiefer}}, \bibinfo {author} {\bibfnamefont
  {S.}~\bibnamefont {Gerischer}}, \bibinfo {author} {\bibfnamefont
  {D.}~\bibnamefont {Slobinsky}}, \ and\ \bibinfo {author} {\bibfnamefont
  {R.~S.}\ \bibnamefont {Perry}},\ }\href@noop {} {\bibfield  {journal}
  {\bibinfo  {journal} {Science}\ }\textbf {\bibinfo {volume} {326}},\ \bibinfo
  {pages} {411} (\bibinfo {year} {2009})}\BibitemShut {NoStop}%
\bibitem [{\citenamefont {Fennell}\ \emph {et~al.}(2009)\citenamefont
  {Fennell}, \citenamefont {Deen}, \citenamefont {Wildes}, \citenamefont
  {Schmalzl}, \citenamefont {Prabhakaran}, \citenamefont {Boothroyd},
  \citenamefont {Aldus}, \citenamefont {McMorrow},\ and\ \citenamefont
  {Bramwell}}]{sc.326.415.2009}%
  \BibitemOpen
  \bibfield  {author} {\bibinfo {author} {\bibfnamefont {T.}~\bibnamefont
  {Fennell}}, \bibinfo {author} {\bibfnamefont {P.~P.}\ \bibnamefont {Deen}},
  \bibinfo {author} {\bibfnamefont {A.~R.}\ \bibnamefont {Wildes}}, \bibinfo
  {author} {\bibfnamefont {K.}~\bibnamefont {Schmalzl}}, \bibinfo {author}
  {\bibfnamefont {D.}~\bibnamefont {Prabhakaran}}, \bibinfo {author}
  {\bibfnamefont {A.~T.}\ \bibnamefont {Boothroyd}}, \bibinfo {author}
  {\bibfnamefont {R.~J.}\ \bibnamefont {Aldus}}, \bibinfo {author}
  {\bibfnamefont {D.~F.}\ \bibnamefont {McMorrow}}, \ and\ \bibinfo {author}
  {\bibfnamefont {S.~T.}\ \bibnamefont {Bramwell}},\ }\href@noop {} {\bibfield
  {journal} {\bibinfo  {journal} {Science}\ }\textbf {\bibinfo {volume}
  {326}},\ \bibinfo {pages} {415} (\bibinfo {year} {2009})}\BibitemShut
  {NoStop}%
\bibitem [{\citenamefont {Gardner}\ \emph {et~al.}(1999)\citenamefont
  {Gardner}, \citenamefont {Dunsiger}, \citenamefont {Gaulin}, \citenamefont
  {Gingras}, \citenamefont {Greedan}, \citenamefont {Kiefl}, \citenamefont
  {Lumsden}, \citenamefont {MacFarlane}, \citenamefont {Raju}, \citenamefont
  {Sonier}, \citenamefont {Swainson},\ and\ \citenamefont
  {Tun}}]{prl.82.1012.1999}%
  \BibitemOpen
  \bibfield  {author} {\bibinfo {author} {\bibfnamefont {J.~S.}\ \bibnamefont
  {Gardner}}, \bibinfo {author} {\bibfnamefont {S.~R.}\ \bibnamefont
  {Dunsiger}}, \bibinfo {author} {\bibfnamefont {B.~D.}\ \bibnamefont
  {Gaulin}}, \bibinfo {author} {\bibfnamefont {M.~J.~P.}\ \bibnamefont
  {Gingras}}, \bibinfo {author} {\bibfnamefont {J.~E.}\ \bibnamefont
  {Greedan}}, \bibinfo {author} {\bibfnamefont {R.~F.}\ \bibnamefont {Kiefl}},
  \bibinfo {author} {\bibfnamefont {M.~D.}\ \bibnamefont {Lumsden}}, \bibinfo
  {author} {\bibfnamefont {W.~A.}\ \bibnamefont {MacFarlane}}, \bibinfo
  {author} {\bibfnamefont {N.~P.}\ \bibnamefont {Raju}}, \bibinfo {author}
  {\bibfnamefont {J.~E.}\ \bibnamefont {Sonier}}, \bibinfo {author}
  {\bibfnamefont {I.}~\bibnamefont {Swainson}}, \ and\ \bibinfo {author}
  {\bibfnamefont {Z.}~\bibnamefont {Tun}},\ }\href@noop {} {\bibfield
  {journal} {\bibinfo  {journal} {Phys. Rev. Lett.}\ }\textbf {\bibinfo
  {volume} {82}},\ \bibinfo {pages} {1012} (\bibinfo {year}
  {1999})}\BibitemShut {NoStop}%
\bibitem [{\citenamefont {Reimers}\ \emph {et~al.}(1991)\citenamefont
  {Reimers}, \citenamefont {Berlinsky},\ and\ \citenamefont
  {Shi}}]{prb.43.865.1991}%
  \BibitemOpen
  \bibfield  {author} {\bibinfo {author} {\bibfnamefont {J.~N.}\ \bibnamefont
  {Reimers}}, \bibinfo {author} {\bibfnamefont {A.~J.}\ \bibnamefont
  {Berlinsky}}, \ and\ \bibinfo {author} {\bibfnamefont {A.-C.}\ \bibnamefont
  {Shi}},\ }\href@noop {} {\bibfield  {journal} {\bibinfo  {journal} {Phys.
  Rev. B}\ }\textbf {\bibinfo {volume} {43}},\ \bibinfo {pages} {865} (\bibinfo
  {year} {1991})}\BibitemShut {NoStop}%
\bibitem [{\citenamefont {Molavian}\ \emph {et~al.}(2007)\citenamefont
  {Molavian}, \citenamefont {Gingras},\ and\ \citenamefont
  {Canals}}]{prl.98.157204.2007}%
  \BibitemOpen
  \bibfield  {author} {\bibinfo {author} {\bibfnamefont {H.~R.}\ \bibnamefont
  {Molavian}}, \bibinfo {author} {\bibfnamefont {M.~J.~P.}\ \bibnamefont
  {Gingras}}, \ and\ \bibinfo {author} {\bibfnamefont {B.}~\bibnamefont
  {Canals}},\ }\href@noop {} {\bibfield  {journal} {\bibinfo  {journal} {Phys.
  Rev. Lett.}\ }\textbf {\bibinfo {volume} {98}},\ \bibinfo {pages} {157204}
  (\bibinfo {year} {2007})}\BibitemShut {NoStop}%
\bibitem [{\citenamefont {Molavian}\ \emph {et~al.}()\citenamefont {Molavian},
  \citenamefont {McClarty},\ and\ \citenamefont
  {Gingras}}]{arxiv.malavian.2009}%
  \BibitemOpen
  \bibfield  {author} {\bibinfo {author} {\bibfnamefont {H.~R.}\ \bibnamefont
  {Molavian}}, \bibinfo {author} {\bibfnamefont {P.~A.}\ \bibnamefont
  {McClarty}}, \ and\ \bibinfo {author} {\bibfnamefont {M.~J.~P.}\ \bibnamefont
  {Gingras}},\ }\href@noop {} {}\Eprint {http://arxiv.org/abs/0912.2957}
  {arXiv:0912.2957} \BibitemShut {NoStop}%
%%CITATION=0912.2957;%%
\bibitem [{\citenamefont {Bonville}\ \emph {et~al.}(2011)\citenamefont
  {Bonville}, \citenamefont {Mirebeau}, \citenamefont {Gukasov}, \citenamefont
  {Petit},\ and\ \citenamefont {Robert}}]{prb.84.184409.2011}%
  \BibitemOpen
  \bibfield  {author} {\bibinfo {author} {\bibfnamefont {P.}~\bibnamefont
  {Bonville}}, \bibinfo {author} {\bibfnamefont {I.}~\bibnamefont {Mirebeau}},
  \bibinfo {author} {\bibfnamefont {A.}~\bibnamefont {Gukasov}}, \bibinfo
  {author} {\bibfnamefont {S.}~\bibnamefont {Petit}}, \ and\ \bibinfo {author}
  {\bibfnamefont {J.}~\bibnamefont {Robert}},\ }\href@noop {} {\bibfield
  {journal} {\bibinfo  {journal} {Phys. Rev. B}\ }\textbf {\bibinfo {volume}
  {84}},\ \bibinfo {pages} {184409} (\bibinfo {year} {2011})}\BibitemShut
  {NoStop}%
\bibitem [{\citenamefont {Cao}\ \emph {et~al.}(2008)\citenamefont {Cao},
  \citenamefont {Gukasov}, \citenamefont {Mirebeau}, \citenamefont {Bonville},\
  and\ \citenamefont {Dhalenne}}]{prl.101.196402.2008}%
  \BibitemOpen
  \bibfield  {author} {\bibinfo {author} {\bibfnamefont {H.}~\bibnamefont
  {Cao}}, \bibinfo {author} {\bibfnamefont {A.}~\bibnamefont {Gukasov}},
  \bibinfo {author} {\bibfnamefont {I.}~\bibnamefont {Mirebeau}}, \bibinfo
  {author} {\bibfnamefont {P.}~\bibnamefont {Bonville}}, \ and\ \bibinfo
  {author} {\bibfnamefont {G.}~\bibnamefont {Dhalenne}},\ }\href@noop {}
  {\bibfield  {journal} {\bibinfo  {journal} {Phys. Rev. Lett.}\ }\textbf
  {\bibinfo {volume} {101}},\ \bibinfo {pages} {196402} (\bibinfo {year}
  {2008})}\BibitemShut {NoStop}%
\bibitem [{\citenamefont {Sazonov}\ \emph {et~al.}(2010)\citenamefont
  {Sazonov}, \citenamefont {Gukasov}, \citenamefont {Mirebeau}, \citenamefont
  {Cao}, \citenamefont {Bonville}, \citenamefont {Grenier},\ and\ \citenamefont
  {Dhalenne}}]{prb.82.174406.2010}%
  \BibitemOpen
  \bibfield  {author} {\bibinfo {author} {\bibfnamefont {A.~P.}\ \bibnamefont
  {Sazonov}}, \bibinfo {author} {\bibfnamefont {A.}~\bibnamefont {Gukasov}},
  \bibinfo {author} {\bibfnamefont {I.}~\bibnamefont {Mirebeau}}, \bibinfo
  {author} {\bibfnamefont {H.}~\bibnamefont {Cao}}, \bibinfo {author}
  {\bibfnamefont {P.}~\bibnamefont {Bonville}}, \bibinfo {author}
  {\bibfnamefont {B.}~\bibnamefont {Grenier}}, \ and\ \bibinfo {author}
  {\bibfnamefont {G.}~\bibnamefont {Dhalenne}},\ }\href@noop {} {\bibfield
  {journal} {\bibinfo  {journal} {Phys. Rev. B}\ }\textbf {\bibinfo {volume}
  {82}},\ \bibinfo {pages} {174406} (\bibinfo {year} {2010})}\BibitemShut
  {NoStop}%
\bibitem [{\citenamefont {Ruff}\ \emph {et~al.}(2005)\citenamefont {Ruff},
  \citenamefont {Melko},\ and\ \citenamefont {Gingras}}]{prl.95.097202.2005}%
  \BibitemOpen
  \bibfield  {author} {\bibinfo {author} {\bibfnamefont {J.~P.~C.}\
  \bibnamefont {Ruff}}, \bibinfo {author} {\bibfnamefont {R.~G.}\ \bibnamefont
  {Melko}}, \ and\ \bibinfo {author} {\bibfnamefont {M.~J.~P.}\ \bibnamefont
  {Gingras}},\ }\href@noop {} {\bibfield  {journal} {\bibinfo  {journal} {Phys.
  Rev. Lett.}\ }\textbf {\bibinfo {volume} {95}},\ \bibinfo {pages} {097202}
  (\bibinfo {year} {2005})}\BibitemShut {NoStop}%
\bibitem [{\citenamefont {Balakrishnan}\ \emph {et~al.}(1998)\citenamefont
  {Balakrishnan}, \citenamefont {Petrenko}, \citenamefont {Lees},\ and\
  \citenamefont {Paul}}]{jpcm.10.L723.1998}%
  \BibitemOpen
  \bibfield  {author} {\bibinfo {author} {\bibfnamefont {G.}~\bibnamefont
  {Balakrishnan}}, \bibinfo {author} {\bibfnamefont {O.~A.}\ \bibnamefont
  {Petrenko}}, \bibinfo {author} {\bibfnamefont {M.~R.}\ \bibnamefont {Lees}},
  \ and\ \bibinfo {author} {\bibfnamefont {D.~M.}\ \bibnamefont {Paul}},\
  }\href@noop {} {\bibfield  {journal} {\bibinfo  {journal} {J. Phys.: Condens.
  Matter}\ }\textbf {\bibinfo {volume} {10}},\ \bibinfo {pages} {L723}
  (\bibinfo {year} {1998})}\BibitemShut {NoStop}%
\bibitem [{\citenamefont {Rodr{\'\i}guez-Carvajal}(1993)}]{phb.192.55.1993}%
  \BibitemOpen
  \bibfield  {author} {\bibinfo {author} {\bibfnamefont {J.}~\bibnamefont
  {Rodr{\'\i}guez-Carvajal}},\ }\href@noop {} {\bibfield  {journal} {\bibinfo
  {journal} {Physica B}\ }\textbf {\bibinfo {volume} {192}},\ \bibinfo {pages}
  {55} (\bibinfo {year} {1993})}\BibitemShut {NoStop}%
\bibitem [{\citenamefont {Rule}\ \emph {et~al.}(2006)\citenamefont {Rule},
  \citenamefont {Ruff}, \citenamefont {Gaulin}, \citenamefont {Dunsiger},
  \citenamefont {Gardner}, \citenamefont {Clancy}, \citenamefont {Lewis},
  \citenamefont {Dabkowska}, \citenamefont {Mirebeau}, \citenamefont {Manuel},
  \citenamefont {Qiu},\ and\ \citenamefont {Copley}}]{prl.96.177201.2006}%
  \BibitemOpen
  \bibfield  {author} {\bibinfo {author} {\bibfnamefont {K.~C.}\ \bibnamefont
  {Rule}}, \bibinfo {author} {\bibfnamefont {J.~P.~C.}\ \bibnamefont {Ruff}},
  \bibinfo {author} {\bibfnamefont {B.~D.}\ \bibnamefont {Gaulin}}, \bibinfo
  {author} {\bibfnamefont {S.~R.}\ \bibnamefont {Dunsiger}}, \bibinfo {author}
  {\bibfnamefont {J.~S.}\ \bibnamefont {Gardner}}, \bibinfo {author}
  {\bibfnamefont {J.~P.}\ \bibnamefont {Clancy}}, \bibinfo {author}
  {\bibfnamefont {M.~J.}\ \bibnamefont {Lewis}}, \bibinfo {author}
  {\bibfnamefont {H.~A.}\ \bibnamefont {Dabkowska}}, \bibinfo {author}
  {\bibfnamefont {I.}~\bibnamefont {Mirebeau}}, \bibinfo {author}
  {\bibfnamefont {P.}~\bibnamefont {Manuel}}, \bibinfo {author} {\bibfnamefont
  {Y.}~\bibnamefont {Qiu}}, \ and\ \bibinfo {author} {\bibfnamefont {J.~R.~D.}\
  \bibnamefont {Copley}},\ }\href@noop {} {\bibfield  {journal} {\bibinfo
  {journal} {Phys. Rev. Lett.}\ }\textbf {\bibinfo {volume} {96}},\ \bibinfo
  {pages} {177201} (\bibinfo {year} {2006})}\BibitemShut {NoStop}%
\bibitem [{\citenamefont {Ruff}\ \emph
  {et~al.}(2010{\natexlab{a}})\citenamefont {Ruff}, \citenamefont {Gaulin},
  \citenamefont {Rule},\ and\ \citenamefont {Gardner}}]{prb.82.100401.2010}%
  \BibitemOpen
  \bibfield  {author} {\bibinfo {author} {\bibfnamefont {J.~P.~C.}\
  \bibnamefont {Ruff}}, \bibinfo {author} {\bibfnamefont {B.~D.}\ \bibnamefont
  {Gaulin}}, \bibinfo {author} {\bibfnamefont {K.~C.}\ \bibnamefont {Rule}}, \
  and\ \bibinfo {author} {\bibfnamefont {J.~S.}\ \bibnamefont {Gardner}},\
  }\href@noop {} {\bibfield  {journal} {\bibinfo  {journal} {Phys. Rev. B}\
  }\textbf {\bibinfo {volume} {82}},\ \bibinfo {pages} {100401} (\bibinfo
  {year} {2010}{\natexlab{a}})}\BibitemShut {NoStop}%
\bibitem [{\citenamefont {Sazonov}\ \emph {et~al.}(2011)\citenamefont
  {Sazonov}, \citenamefont {Gukasov},\ and\ \citenamefont
  {Mirebeau}}]{jpcm.23.164221.2011}%
  \BibitemOpen
  \bibfield  {author} {\bibinfo {author} {\bibfnamefont {A.~P.}\ \bibnamefont
  {Sazonov}}, \bibinfo {author} {\bibfnamefont {A.}~\bibnamefont {Gukasov}}, \
  and\ \bibinfo {author} {\bibfnamefont {I.}~\bibnamefont {Mirebeau}},\
  }\href@noop {} {\bibfield  {journal} {\bibinfo  {journal} {J. Phys.: Condens.
  Matter}\ }\textbf {\bibinfo {volume} {23}},\ \bibinfo {pages} {164221}
  (\bibinfo {year} {2011})}\BibitemShut {NoStop}%
\bibitem [{\citenamefont {Yasui}\ \emph {et~al.}(2002)\citenamefont {Yasui},
  \citenamefont {Kanada}, \citenamefont {Ito}, \citenamefont {Harashina},
  \citenamefont {Sato}, \citenamefont {Okumara}, \citenamefont {Kakurai},\ and\
  \citenamefont {Kadowaki}}]{jpsj.71.599.2002}%
  \BibitemOpen
  \bibfield  {author} {\bibinfo {author} {\bibfnamefont {Y.}~\bibnamefont
  {Yasui}}, \bibinfo {author} {\bibfnamefont {M.}~\bibnamefont {Kanada}},
  \bibinfo {author} {\bibfnamefont {M.}~\bibnamefont {Ito}}, \bibinfo {author}
  {\bibfnamefont {H.}~\bibnamefont {Harashina}}, \bibinfo {author}
  {\bibfnamefont {M.}~\bibnamefont {Sato}}, \bibinfo {author} {\bibfnamefont
  {H.}~\bibnamefont {Okumara}}, \bibinfo {author} {\bibfnamefont
  {K.}~\bibnamefont {Kakurai}}, \ and\ \bibinfo {author} {\bibfnamefont
  {H.}~\bibnamefont {Kadowaki}},\ }\href@noop {} {\bibfield  {journal}
  {\bibinfo  {journal} {J. Phys. Soc. Jpn.}\ }\textbf {\bibinfo {volume}
  {71}},\ \bibinfo {pages} {599} (\bibinfo {year} {2002})}\BibitemShut
  {NoStop}%
\bibitem [{\citenamefont {Kadowaki}\ \emph {et~al.}(2009)\citenamefont
  {Kadowaki}, \citenamefont {Doi}, \citenamefont {Aoki}, \citenamefont
  {Tabata}, \citenamefont {Sato}, \citenamefont {Lynn}, \citenamefont
  {Matsuhira},\ and\ \citenamefont {Hiroi}}]{jpsj.78.103706.2009}%
  \BibitemOpen
  \bibfield  {author} {\bibinfo {author} {\bibfnamefont {H.}~\bibnamefont
  {Kadowaki}}, \bibinfo {author} {\bibfnamefont {N.}~\bibnamefont {Doi}},
  \bibinfo {author} {\bibfnamefont {Y.}~\bibnamefont {Aoki}}, \bibinfo {author}
  {\bibfnamefont {Y.}~\bibnamefont {Tabata}}, \bibinfo {author} {\bibfnamefont
  {T.~J.}\ \bibnamefont {Sato}}, \bibinfo {author} {\bibfnamefont {J.~W.}\
  \bibnamefont {Lynn}}, \bibinfo {author} {\bibfnamefont {K.}~\bibnamefont
  {Matsuhira}}, \ and\ \bibinfo {author} {\bibfnamefont {Z.}~\bibnamefont
  {Hiroi}},\ }\href@noop {} {\bibfield  {journal} {\bibinfo  {journal} {J.
  Phys. Soc. Jpn.}\ }\textbf {\bibinfo {volume} {78}},\ \bibinfo {pages}
  {103706} (\bibinfo {year} {2009})}\BibitemShut {NoStop}%
\bibitem [{\citenamefont {Harris}\ \emph {et~al.}(1997)\citenamefont {Harris},
  \citenamefont {Bramwell}, \citenamefont {McMorrow}, \citenamefont {Zeiske},\
  and\ \citenamefont {Godfrey}}]{prl.79.2554.1997}%
  \BibitemOpen
  \bibfield  {author} {\bibinfo {author} {\bibfnamefont {M.~J.}\ \bibnamefont
  {Harris}}, \bibinfo {author} {\bibfnamefont {S.~T.}\ \bibnamefont
  {Bramwell}}, \bibinfo {author} {\bibfnamefont {D.~F.}\ \bibnamefont
  {McMorrow}}, \bibinfo {author} {\bibfnamefont {T.}~\bibnamefont {Zeiske}}, \
  and\ \bibinfo {author} {\bibfnamefont {K.~W.}\ \bibnamefont {Godfrey}},\
  }\href@noop {} {\bibfield  {journal} {\bibinfo  {journal} {Phys. Rev. Lett.}\
  }\textbf {\bibinfo {volume} {79}},\ \bibinfo {pages} {2554} (\bibinfo {year}
  {1997})}\BibitemShut {NoStop}%
\bibitem [{\citenamefont {Melko}\ and\ \citenamefont
  {Gingras}(2004)}]{jpcm.16.R1277.2004}%
  \BibitemOpen
  \bibfield  {author} {\bibinfo {author} {\bibfnamefont {R.~G.}\ \bibnamefont
  {Melko}}\ and\ \bibinfo {author} {\bibfnamefont {M.~J.~P.}\ \bibnamefont
  {Gingras}},\ }\href@noop {} {\bibfield  {journal} {\bibinfo  {journal} {J.
  Phys.: Condens. Matter}\ }\textbf {\bibinfo {volume} {16}},\ \bibinfo {pages}
  {R1277} (\bibinfo {year} {2004})}\BibitemShut {NoStop}%
\bibitem [{\citenamefont {Clancy}\ \emph {et~al.}(2009)\citenamefont {Clancy},
  \citenamefont {Ruff}, \citenamefont {Dunsiger}, \citenamefont {Zhao},
  \citenamefont {Dabkowska}, \citenamefont {Gardner}, \citenamefont {Qiu},
  \citenamefont {Copley}, \citenamefont {Jenkins},\ and\ \citenamefont
  {Gaulin}}]{prb.79.014408.2009}%
  \BibitemOpen
  \bibfield  {author} {\bibinfo {author} {\bibfnamefont {J.~P.}\ \bibnamefont
  {Clancy}}, \bibinfo {author} {\bibfnamefont {J.~P.~C.}\ \bibnamefont {Ruff}},
  \bibinfo {author} {\bibfnamefont {S.~R.}\ \bibnamefont {Dunsiger}}, \bibinfo
  {author} {\bibfnamefont {Y.}~\bibnamefont {Zhao}}, \bibinfo {author}
  {\bibfnamefont {H.~A.}\ \bibnamefont {Dabkowska}}, \bibinfo {author}
  {\bibfnamefont {J.~S.}\ \bibnamefont {Gardner}}, \bibinfo {author}
  {\bibfnamefont {Y.}~\bibnamefont {Qiu}}, \bibinfo {author} {\bibfnamefont
  {J.~R.~D.}\ \bibnamefont {Copley}}, \bibinfo {author} {\bibfnamefont
  {T.}~\bibnamefont {Jenkins}}, \ and\ \bibinfo {author} {\bibfnamefont
  {B.~D.}\ \bibnamefont {Gaulin}},\ }\href@noop {} {\bibfield  {journal}
  {\bibinfo  {journal} {Phys. Rev. B}\ }\textbf {\bibinfo {volume} {79}},\
  \bibinfo {pages} {014408} (\bibinfo {year} {2009})}\BibitemShut {NoStop}%
\bibitem [{\citenamefont {Mirebeau}\ \emph {et~al.}(2007)\citenamefont
  {Mirebeau}, \citenamefont {Bonville},\ and\ \citenamefont
  {Hennion}}]{prb.76.184436.2007}%
  \BibitemOpen
  \bibfield  {author} {\bibinfo {author} {\bibfnamefont {I.}~\bibnamefont
  {Mirebeau}}, \bibinfo {author} {\bibfnamefont {P.}~\bibnamefont {Bonville}},
  \ and\ \bibinfo {author} {\bibfnamefont {M.}~\bibnamefont {Hennion}},\
  }\href@noop {} {\bibfield  {journal} {\bibinfo  {journal} {Phys. Rev. B}\
  }\textbf {\bibinfo {volume} {76}},\ \bibinfo {pages} {184436} (\bibinfo
  {year} {2007})}\BibitemShut {NoStop}%
\bibitem [{\citenamefont {Cao}\ \emph {et~al.}(2009)\citenamefont {Cao},
  \citenamefont {Gukasov}, \citenamefont {Mirebeau}, \citenamefont {Bonville},
  \citenamefont {Decorse},\ and\ \citenamefont
  {Dhalenne}}]{prl.103.056402.2009}%
  \BibitemOpen
  \bibfield  {author} {\bibinfo {author} {\bibfnamefont {H.}~\bibnamefont
  {Cao}}, \bibinfo {author} {\bibfnamefont {A.}~\bibnamefont {Gukasov}},
  \bibinfo {author} {\bibfnamefont {I.}~\bibnamefont {Mirebeau}}, \bibinfo
  {author} {\bibfnamefont {P.}~\bibnamefont {Bonville}}, \bibinfo {author}
  {\bibfnamefont {C.}~\bibnamefont {Decorse}}, \ and\ \bibinfo {author}
  {\bibfnamefont {G.}~\bibnamefont {Dhalenne}},\ }\href@noop {} {\bibfield
  {journal} {\bibinfo  {journal} {Phys. Rev. Lett.}\ }\textbf {\bibinfo
  {volume} {103}},\ \bibinfo {pages} {056402} (\bibinfo {year}
  {2009})}\BibitemShut {NoStop}%
\bibitem [{\citenamefont {Hiroi}\ \emph {et~al.}(2003)\citenamefont {Hiroi},
  \citenamefont {Matsuhira},\ and\ \citenamefont {Ogata}}]{jpsj.72.3045.2003}%
  \BibitemOpen
  \bibfield  {author} {\bibinfo {author} {\bibfnamefont {Z.}~\bibnamefont
  {Hiroi}}, \bibinfo {author} {\bibfnamefont {K.}~\bibnamefont {Matsuhira}}, \
  and\ \bibinfo {author} {\bibfnamefont {M.}~\bibnamefont {Ogata}},\
  }\href@noop {} {\bibfield  {journal} {\bibinfo  {journal} {J. Phys. Soc.
  Jpn.}\ }\textbf {\bibinfo {volume} {72}},\ \bibinfo {pages} {3045} (\bibinfo
  {year} {2003})}\BibitemShut {NoStop}%
\bibitem [{\citenamefont {Ruff}\ \emph {et~al.}(2007)\citenamefont {Ruff},
  \citenamefont {Gaulin}, \citenamefont {Castellan}, \citenamefont {Rule},
  \citenamefont {Clancy}, \citenamefont {Rodriguez},\ and\ \citenamefont
  {Dabkowska}}]{prl.99.237202.2007}%
  \BibitemOpen
  \bibfield  {author} {\bibinfo {author} {\bibfnamefont {J.~P.~C.}\
  \bibnamefont {Ruff}}, \bibinfo {author} {\bibfnamefont {B.~D.}\ \bibnamefont
  {Gaulin}}, \bibinfo {author} {\bibfnamefont {J.~P.}\ \bibnamefont
  {Castellan}}, \bibinfo {author} {\bibfnamefont {K.~C.}\ \bibnamefont {Rule}},
  \bibinfo {author} {\bibfnamefont {J.~P.}\ \bibnamefont {Clancy}}, \bibinfo
  {author} {\bibfnamefont {J.}~\bibnamefont {Rodriguez}}, \ and\ \bibinfo
  {author} {\bibfnamefont {H.~A.}\ \bibnamefont {Dabkowska}},\ }\href@noop {}
  {\bibfield  {journal} {\bibinfo  {journal} {Phys. Rev. Lett.}\ }\textbf
  {\bibinfo {volume} {99}},\ \bibinfo {pages} {237202} (\bibinfo {year}
  {2007})}\BibitemShut {NoStop}%
\bibitem [{\citenamefont {Yamashita}\ and\ \citenamefont
  {Ueda}(2000)}]{prl.85.4960.2000}%
  \BibitemOpen
  \bibfield  {author} {\bibinfo {author} {\bibfnamefont {Y.}~\bibnamefont
  {Yamashita}}\ and\ \bibinfo {author} {\bibfnamefont {K.}~\bibnamefont
  {Ueda}},\ }\href@noop {} {\bibfield  {journal} {\bibinfo  {journal} {Phys.
  Rev. Lett.}\ }\textbf {\bibinfo {volume} {85}},\ \bibinfo {pages} {4960}
  (\bibinfo {year} {2000})}\BibitemShut {NoStop}%
\bibitem [{\citenamefont {Tchernyshyov}\ \emph {et~al.}(2002)\citenamefont
  {Tchernyshyov}, \citenamefont {Moessner},\ and\ \citenamefont
  {Sondhi}}]{prl.88.067203.2002}%
  \BibitemOpen
  \bibfield  {author} {\bibinfo {author} {\bibfnamefont {O.}~\bibnamefont
  {Tchernyshyov}}, \bibinfo {author} {\bibfnamefont {R.}~\bibnamefont
  {Moessner}}, \ and\ \bibinfo {author} {\bibfnamefont {S.~L.}\ \bibnamefont
  {Sondhi}},\ }\href@noop {} {\bibfield  {journal} {\bibinfo  {journal} {Phys.
  Rev. Lett.}\ }\textbf {\bibinfo {volume} {88}},\ \bibinfo {pages} {067203}
  (\bibinfo {year} {2002})}\BibitemShut {NoStop}%
\bibitem [{\citenamefont {Ruff}\ \emph
  {et~al.}(2010{\natexlab{b}})\citenamefont {Ruff}, \citenamefont {Islam},
  \citenamefont {Clancy}, \citenamefont {Ross}, \citenamefont {Nojiri},
  \citenamefont {Matsuda}, \citenamefont {Dabkowska}, \citenamefont
  {Dabkowski},\ and\ \citenamefont {Gaulin}}]{prl.105.077203.2010}%
  \BibitemOpen
  \bibfield  {author} {\bibinfo {author} {\bibfnamefont {J.~P.~C.}\
  \bibnamefont {Ruff}}, \bibinfo {author} {\bibfnamefont {Z.}~\bibnamefont
  {Islam}}, \bibinfo {author} {\bibfnamefont {J.~P.}\ \bibnamefont {Clancy}},
  \bibinfo {author} {\bibfnamefont {K.~A.}\ \bibnamefont {Ross}}, \bibinfo
  {author} {\bibfnamefont {H.}~\bibnamefont {Nojiri}}, \bibinfo {author}
  {\bibfnamefont {Y.~H.}\ \bibnamefont {Matsuda}}, \bibinfo {author}
  {\bibfnamefont {H.~A.}\ \bibnamefont {Dabkowska}}, \bibinfo {author}
  {\bibfnamefont {A.~D.}\ \bibnamefont {Dabkowski}}, \ and\ \bibinfo {author}
  {\bibfnamefont {B.~D.}\ \bibnamefont {Gaulin}},\ }\href@noop {} {\bibfield
  {journal} {\bibinfo  {journal} {Phys. Rev. Lett.}\ }\textbf {\bibinfo
  {volume} {105}},\ \bibinfo {pages} {077203} (\bibinfo {year}
  {2010}{\natexlab{b}})}\BibitemShut {NoStop}%
\bibitem [{\citenamefont {Mirebeau}\ \emph {et~al.}(2004)\citenamefont
  {Mirebeau}, \citenamefont {Goncharenko}, \citenamefont {Dhalenne},\ and\
  \citenamefont {Revcolevschi}}]{prl.93.187204.2004}%
  \BibitemOpen
  \bibfield  {author} {\bibinfo {author} {\bibfnamefont {I.}~\bibnamefont
  {Mirebeau}}, \bibinfo {author} {\bibfnamefont {I.~N.}\ \bibnamefont
  {Goncharenko}}, \bibinfo {author} {\bibfnamefont {G.}~\bibnamefont
  {Dhalenne}}, \ and\ \bibinfo {author} {\bibfnamefont {A.}~\bibnamefont
  {Revcolevschi}},\ }\href@noop {} {\bibfield  {journal} {\bibinfo  {journal}
  {Phys. Rev. Lett.}\ }\textbf {\bibinfo {volume} {93}},\ \bibinfo {pages}
  {187204} (\bibinfo {year} {2004})}\BibitemShut {NoStop}%
\end{thebibliography}
%\end{document}

%%%%%%%%%%%%%%%%%%%%%%%%%%%%%%%%%%%%%%%%%%%%%%%%%%%%%%%%%%%%%%%%%%%%%%%%%%%%%%%
%%%%%%%%%%%%%%%%%%%%%%%%%%%%%%%%%%%%%%%%%%%%%%%%%%%%%%%%%%%%%%%%%%%%%%%%%%%%%%%
%%%%%%%%%%%%%%%%%%%%%%%%%%%%%%%%%%%%%%%%%%%%%%%%%%%%%%%%%%%%%%%%%%%%%%%%%%%%%%%
% copy of bbl file
% ----------
%merlin.mbs apsrev4-1.bst 2010-07-25 4.21a (PWD, AO, DPC) hacked
%Control: key (0)
%Control: author (8) initials jnrlst
%Control: editor formatted (1) identically to author
%Control: production of article title (-1) disabled
%Control: page (0) single
%Control: year (1) truncated
%Control: production of eprint (0) enabled
%
% ----------

\end{document}